\documentclass[conf]{new-aiaa}
\usepackage[utf8]{inputenc}
\usepackage{graphicx}
\usepackage{amsmath}
\usepackage{amsfonts}
\usepackage{makeidx}
\usepackage{booktabs}
\usepackage{tabularx}
\usepackage{caption}
\usepackage{textcomp}
\usepackage{caption}
\usepackage{subcaption}
\usepackage[version=4]{mhchem}
\usepackage{siunitx}
\usepackage{longtable,tabularx}
\usepackage{array}
\usepackage{hyperref}
\usepackage{float}
\usepackage{atbegshi,picture}
\usepackage{lipsum}
\usepackage[export]{adjustbox}
\title{LES/DNS of flow past T106 LPT cascade using a higher-order LB model}
\author{N. H. Maruthi\footnote{Technical Lead, Aerospace and Automotive group, AIAA Member, Email address: maruthinh@sankhyasutralabs.com} 
}
\author{Chakradhar Thantanapally \footnote{Technical Manager, Aerospace and Automotive group, AIAA Member, Email address: chakradhar@sankhyasutralabs.com}
}
\affil{SankhyaSutra Labs Ltd., Bengaluru, Karnataka, India, 560047}
\author{Manjusha Namburi \footnote{Former PhD student, Email address: manjusha.namburi88@gmail.com}
}
\affil{Jawaharlal Nehru Centre for Advanced Scientific Research (JNCASR), Bengaluru, Karnataka, India, 560064}
\author{V. Kumaran \footnote{Technical Advisor, also professor at the Indian Institute of Science, Bengaluru, Karnataka, India, 560012, Email address: kumaran@sankhyasutralabs.com}}
\author{Santosh Ansumali\footnote{Chief Technology Officer (CTO), also professor at the Jawaharlal Nehru Centre for Advanced Scientific Research (JNCASR), Bengaluru, Karnataka, India, 560064, Email address: ansumali@sankhyasutralabs.com}}
\affil{SankhyaSutra Labs Ltd., Bengaluru, Karnataka, India, 560047}
\begin{document}
\maketitle
\begin{abstract}

	The main objective of the present work is to assess higher-order Entropic Lattice Boltzmann Method (ELBM) for separated and transitional flows without the use of any explicit turbulence model. For this, we chose to simulate two cases of T106 Low-Pressure Turbine (LPT) cascade \textemdash T106A and T106C \textemdash representing incompressible and compressible flow regimes respectively. These results are obtained using our company's in-house higher-order ELBM transonic solver. We have carried out two sets of simulations for both the test cases. One with a clean inlet and the other with an inlet disturbance given by white Gaussian noise superimposed on the inlet velocity. For the clean inlet case, the flow remains laminar on the entire blade surface for both the test cases. It undergoes transition on the suction side for the inlet disturbance case. The pressure coefficient for the T106A and the isentropic Mach number on the blade surface for the T106C matches well with the experimental results.  Also, the qualitative comparison of flow features in terms of the Q-criterion is in good agreement with the earlier computational results reported in literature.
\end{abstract}
%
\section{Introduction}
The flow through a gas turbine is known to be very complicated and, in particular, on the blades it is often called a flow zoo. This is because it exhibits a rich diversity of flow phenomena such as separation,
transition, relaminarization and effects of the high curvature of the blade on the boundary layer,
heat transfer, etc. RANS models which are routinely used in industry for design are inadequate in accurately
predicting these features in turbulent flows. Because of these and other reasons, the gas turbine engine
is known to be one of the most complex pieces of mechanical engineering. An accurate understanding of these
flow features requires DNS/LES like simulations on a well-resolved mesh.
To carry out these simulations, we use our in-house Entropic Lattice Boltzmann Method (ELBM) solver. The details of the model are tabulated in Table~\ref{table:table3}. LBM is a mesoscopic method that solves the Boltzmann equation using an appropriate probability distribution function. Macroscopic properties such as density, velocity, pressure, temperature, etc. are derived as moments of the distribution function. LBM is becoming an increasingly popular alternative to traditional CFD methods owing to its simplicity and scalability on parallel computing architectures \cite{succi2015lattice, succi2001lattice}. However, the standard LB methods were unstable due to the lack of the discrete H-theorem, especially when the mesh is under- resolved \cite{karlin1999perfect, succi2002colloquium}. A class of new LB methods called Entropic Lattice Boltzmann Models (ELBM) are developed to overcome these stability issues \cite{karlin1998maximum, ansumali2002entropy, ansumali2003minimal, chikatamarla2006entropic, atif2017essentially}. The recently introduced crystallographic LBM further improves the capability of LBM by optimally sampling the domain on a Body-Centred Cubic (BCC) lattice instead of the standard Simple Cubic (SC) lattice \cite{namburi2016crystallographic}.
The BCC lattice also improves the discretization of geometries in the computational domain. Furthermore, higher-order LB methods have been introduced to extend the applicability  over a wider range of Mach and Knudsen numbers. These higher-order models are characterized by improved stability, accuracy, and thermal properties \cite{kolluru2020lattice, atif2018higher}.
The use of ELBM ensures a seamless transition from LES to DNS based only on the increasing the grid
resolution without any empirical modelling for small scales. This ensures that the flow features near the
wall are accurately resolved by performing DNS near the wall when the fine resolution is used. Under-resolved simulations give results similar to that of LES.
\begin{table}\centering
	\begin{tabular}{| m{2cm} |c | m{3cm} |}
		\hline
		Shells             & Discrete Velocities($c_i$)                                                                                          & Weight($w_i$)           \\ \hline
		SC - 1             & $\left(\pm 1, 0, 0 \right), \left( 0, \pm1, 0\right), \left(0, 0, \pm1 \right)$                                     & 0.0111443971            \\
		SC - 2             & $\left(\pm 2, 0, 0 \right), \left( 0, \pm2, 0\right), \left(0, 0, \pm2 \right)$                                     & 0.0009925820            \\
		SC - 3             & $\left(\pm 3, 0, 0 \right), \left( 0, \pm3, 0\right), \left(0, 0, \pm3 \right)$                                     & 0.0004322964            \\
		FCC-1              & $ \left(\pm1, \pm1, 0 \right), \left( 0, \pm1, \pm1\right), \left( \pm1, \pm1, 0\right) $                           & 0.0078020595            \\
		FCC-2              & $ \left(\pm2, \pm2, 0 \right), \left( 0, \pm2 , \pm2\right), \left( \pm2 , \pm2, 0\right) $                         & 0.0008848260            \\
		FCC-3              & $ \left(\pm3, \pm3, 0 \right), \left( 0, \pm3 , \pm3\right), \left( \pm3 , \pm3, 0\right) $                         & $6.6974 \times 10^{-6}$ \\
		BCC-1              & $ \left( \pm 1, \pm1, \pm1\right)$                                                                                  & 0.0071479427            \\
		BCC-2              & $ \left( \pm 2, \pm2, \pm2\right)$                                                                                  & 0.0000121756            \\
		BCC-h              & $\left(\pm 1/2, \pm 1/2, \pm 1/2\right)$                                                                            & 0.0447801239            \\
		BCC-3h             & $\left(\pm 3/2, \pm 3/2, \pm 3/2 \right)$                                                                           & 0.0026467301            \\
		BCC-5h             & $\left(\pm 5/2, \pm 5/2, \pm 5/2 \right)$                                                                           & $4.8164 \times 10^{-6}$ \\
		offdiagonal BCC- h & $ \left( \pm 1/2, \pm1, \pm1\right),$ $ \left( \pm 1, \pm 1/2, \pm1\right)$, $ \left( \pm 1, \pm1, \pm 1/2 \right)$ & 0.0103828723            \\
		offdiagonal BCC-1  & $ \left( \pm 2, \pm1, \pm1\right),$ $ \left( \pm 1, \pm2, \pm1\right)$, $ \left( \pm 1, \pm1, \pm2\right)$          & 0.0002149798            \\
		offdiagonal FCC    & $ \left(\pm2, \pm1, 0 \right), \left( 0, \pm2, \pm1\right), \left( \pm2, \pm1, 0\right) $                           &                         \\
		                   & $ \left(\pm1, \pm2, 0 \right), \left( 0, \pm1, \pm2\right), \left( \pm1, \pm2, 0\right) $                           & 0.0041472523            \\ \hline
	\end{tabular}
	\caption{Energy shells and their corresponding velocities with weights for the RD3Q167 model.}
	\label{table:table3}
\end{table}
\section{Computational setup}
\subsection{T106A}
\begin{figure}[h!]
	\centering
	\includegraphics[width=0.75\linewidth]{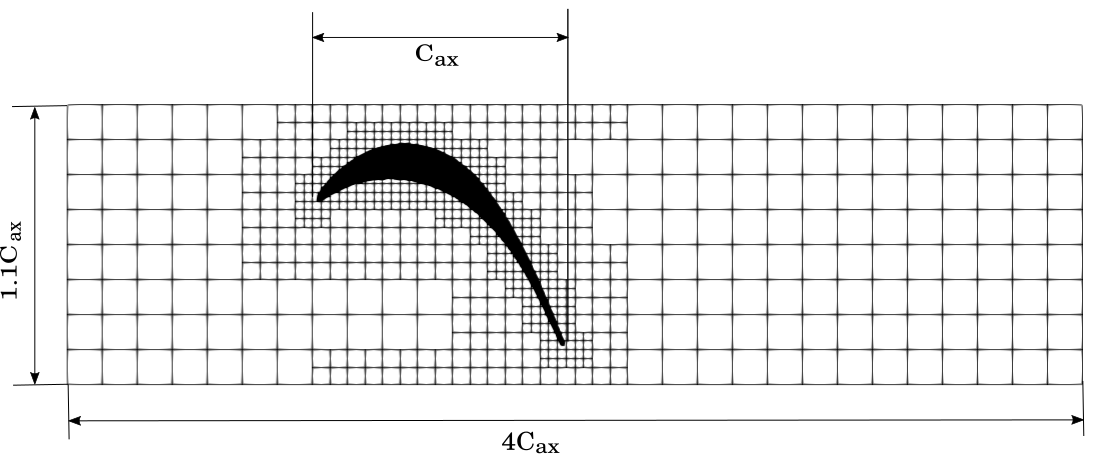}
	\caption{Computational setup used in the simulations is shown along with distribution of grid resolution with the finest near the surface to accurately capture the boundary layer}
	\label{fig:comp_setup}
\end{figure}

The Low-Pressure Turbine (LPT) blade geometry (T106) chosen for simulation is of the Pratt and Whitney rotor named PW2037. The blade has a chord length of 198mm, an axial chord length of 170mm, and a pitch of 158mm \cite{stieger2002effect}. The computational setup consists of one T106 blade with periodicity in a pitch-wise direction to represent the turbine passage. The inflow boundary is located at $C_{ax}$ (axial chord length) upstream from the leading edge, and the outflow boundary is located at $2C_{ax}$ downstream from the trailing edge as shown in Fig. \ref{fig:comp_setup}. The computational domain's spanwise extent is chosen as 20\% of the axial chord length. Uniform inflow and constant outflow boundary conditions are used for the inlet and outlet. Periodic boundary conditions are used in cross-stream and spanwise directions. The diffuse bounce back method is applied on the blade surface \cite{PhysRevE.89.033313}. The multi-resolution approach with a finer resolution near the blade surface and coarser-resolution away from the blade reduces the computational cost (see Fig \ref{fig:comp_setup}). Flow parameters chosen for simulations are similar to those of Kalitzin et al. \cite{kalitzin2003dns}. The inflow Mach number is $0.1$, the inlet flow angle is $37.7^o$, and the Reynolds number based on axial chord length is $148,000$.
\subsection{T106C}
This is one of the test cases chosen in International Workshop on High-Order CFD Methods (HiOCFD) to assess the CFD solvers for their ability to resolve separated and transitional flows\cite{KoenHillewaert}. This is a high-lift blade with higher blade loading, leading to large separation on the suction side \cite{michakalek2012aerodynamic}. This blade has a chord length of 93.01mm, an axial chord of 79.9mm and a pitch of 88.36mm. The computational domain and setup for T106C is similar to that of T106A with the same domain length in all the three directions. Total pressure and temperature based subsonic inflow boundary condition and static pressure outlet is prescribed. Periodic boundary conditions are specified for the cross-stream and pitch wise directions. For this test case, exit Reynolds number based on chord length is $80,000$, Mach number is $0.65$ and inflow angle is $32.7^o.$
\section{Results and Discussion}
\subsection{T106A}
\begin{figure}[h!]
	\centering
	\begin{subfigure}[b]{0.49\linewidth}
		\includegraphics[width=\linewidth]{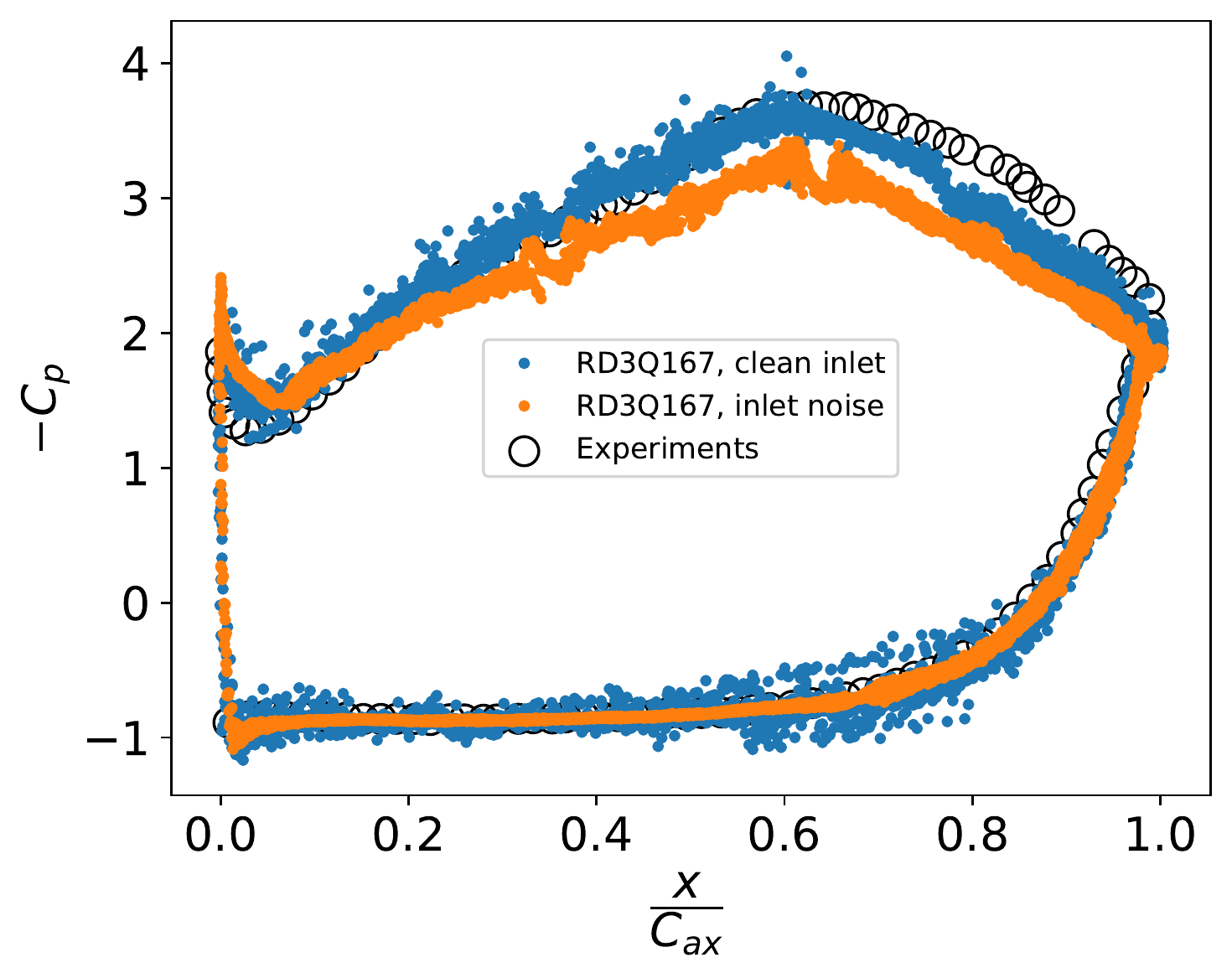}
	\end{subfigure}
	\caption{Averaged Pressure coefficient ($C_p$) on the surface of blade at its mid-span}
	\label{fig:cp_t106a}
\end{figure}

\begin{figure}[h!]
	\centering
	\begin{subfigure}[b]{0.4\linewidth}
		\includegraphics[width=\linewidth]{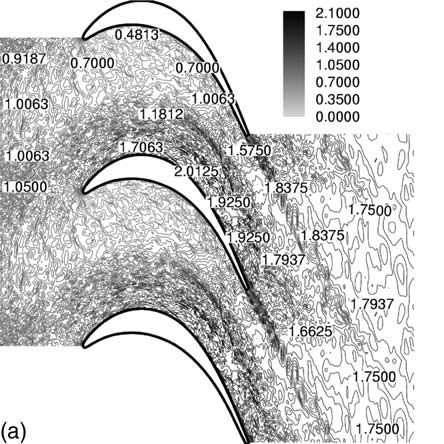}
	\end{subfigure}
	\begin{subfigure}[b]{0.59\linewidth}
		\includegraphics[width=\linewidth]{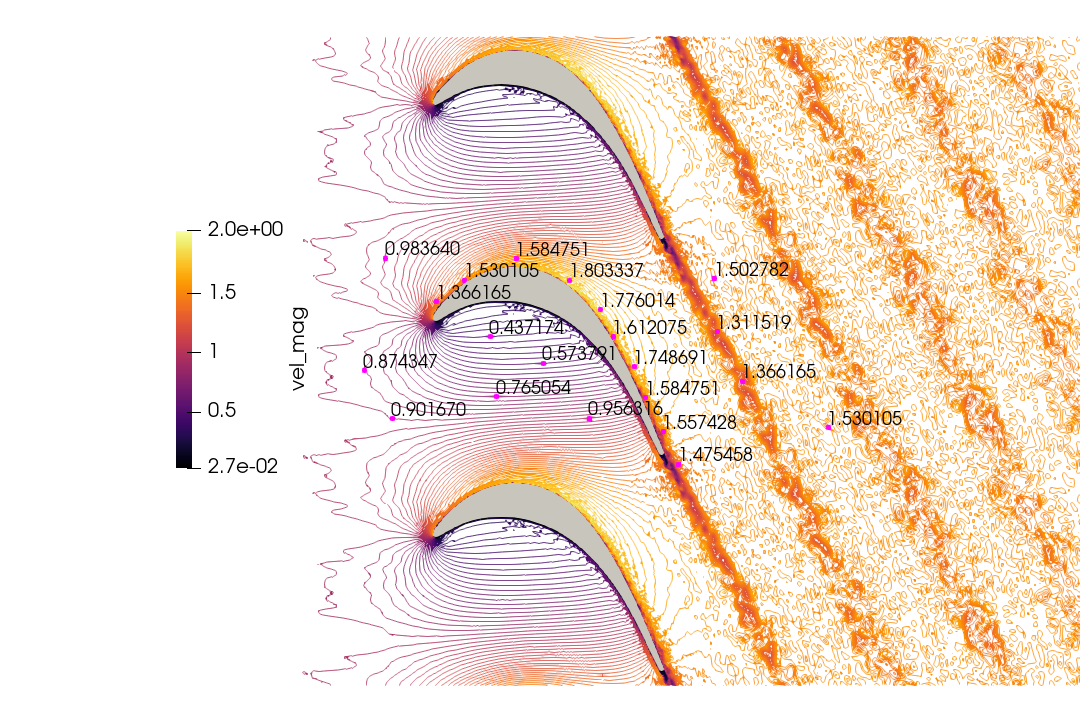}
	\end{subfigure}
	\caption{Instantaneous velocity magnitude (with inlet disturbance (b)) in comparison with reference solution (a) which is obtained with grid turbulence.}
	\label{fig:inst_mag}
\end{figure}

\begin{figure}[h!]
	\centering
	\begin{subfigure}[b]{0.45\linewidth}
		\includegraphics[width=\linewidth]{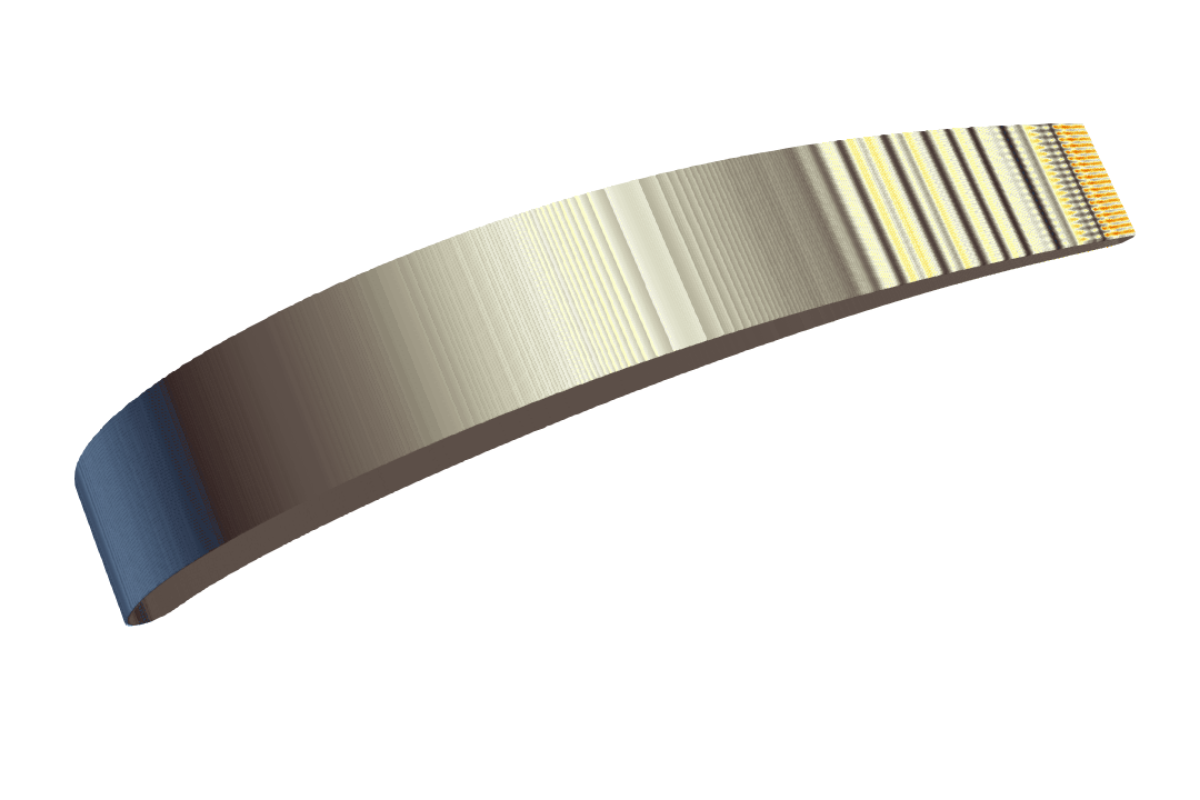}
		\caption{}
	\end{subfigure}
	\begin{subfigure}[b]{0.45\linewidth}
		\includegraphics[width=\linewidth]{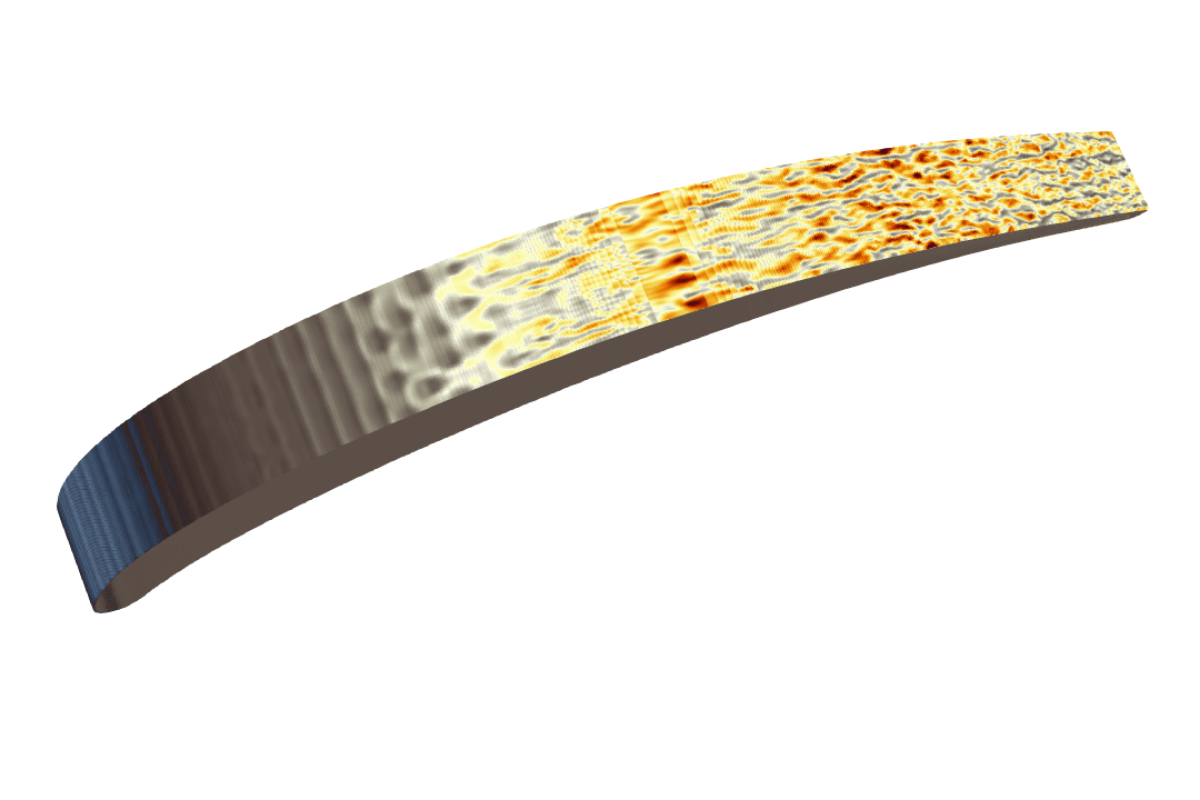}
		\caption{}
	\end{subfigure}
	\begin{subfigure}[b]{0.6\linewidth}
		\includegraphics[width=\linewidth]{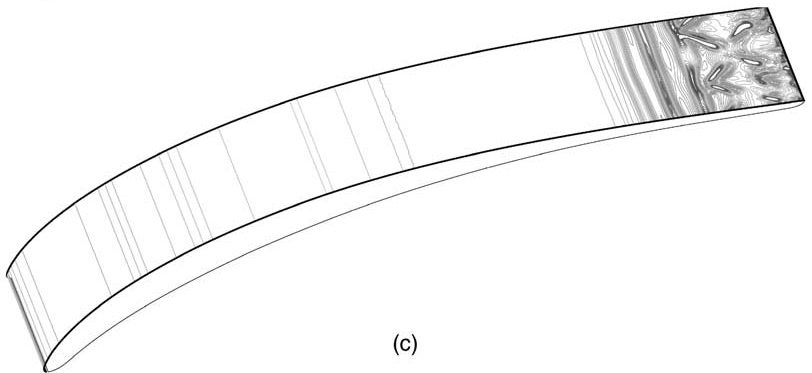}
	\end{subfigure}
	\caption{ The v component of instantaneous velocity on the blade surface to visualise transition. (a) clean inlet, (b) with inlet disturbance and (c) velocity normal to surface (clean inlet)}
	\label{fig:trans}
\end{figure}

\begin{figure}[h!]
	\centering
	\begin{subfigure}[b]{0.49\linewidth}
		\includegraphics[trim = 0cm 0cm 0cm 0cm, clip, width=\linewidth]{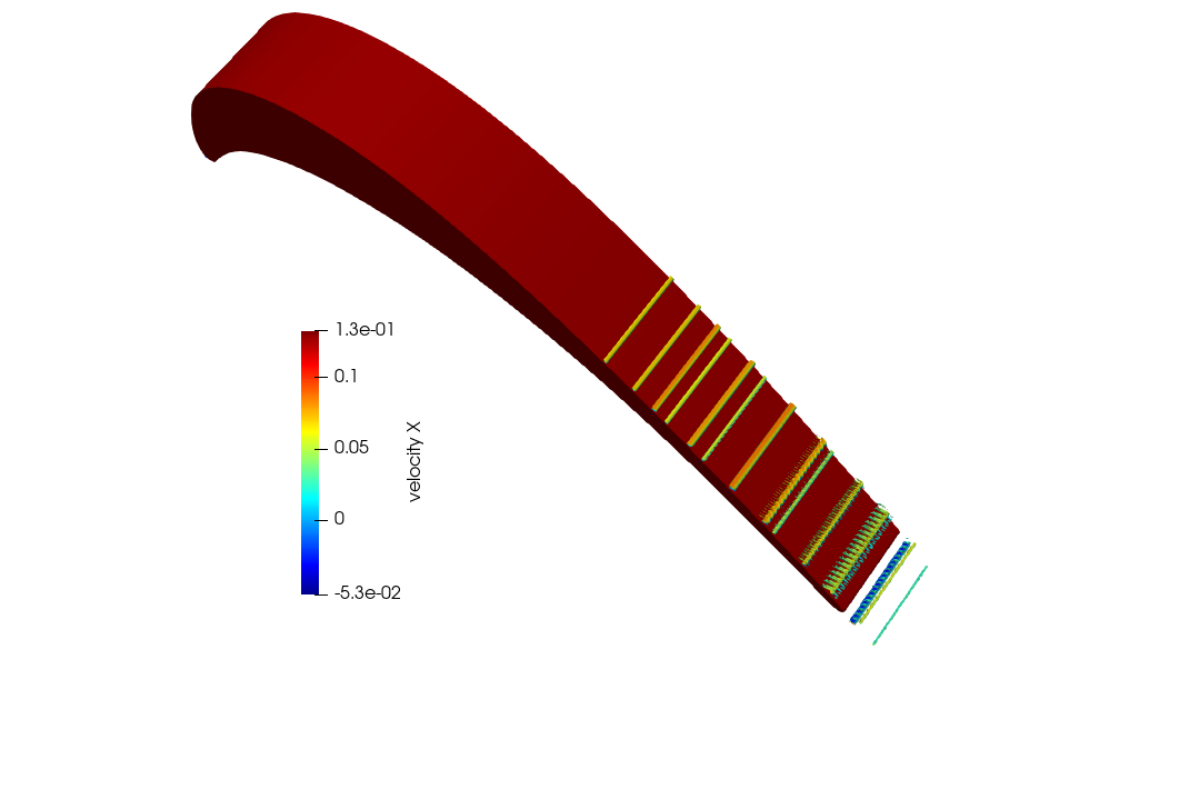}
		\caption{}
	\end{subfigure}
	\begin{subfigure}[b]{0.49\linewidth}
		\includegraphics[trim = 0cm 0cm 0cm 0cm, clip, width=\linewidth]{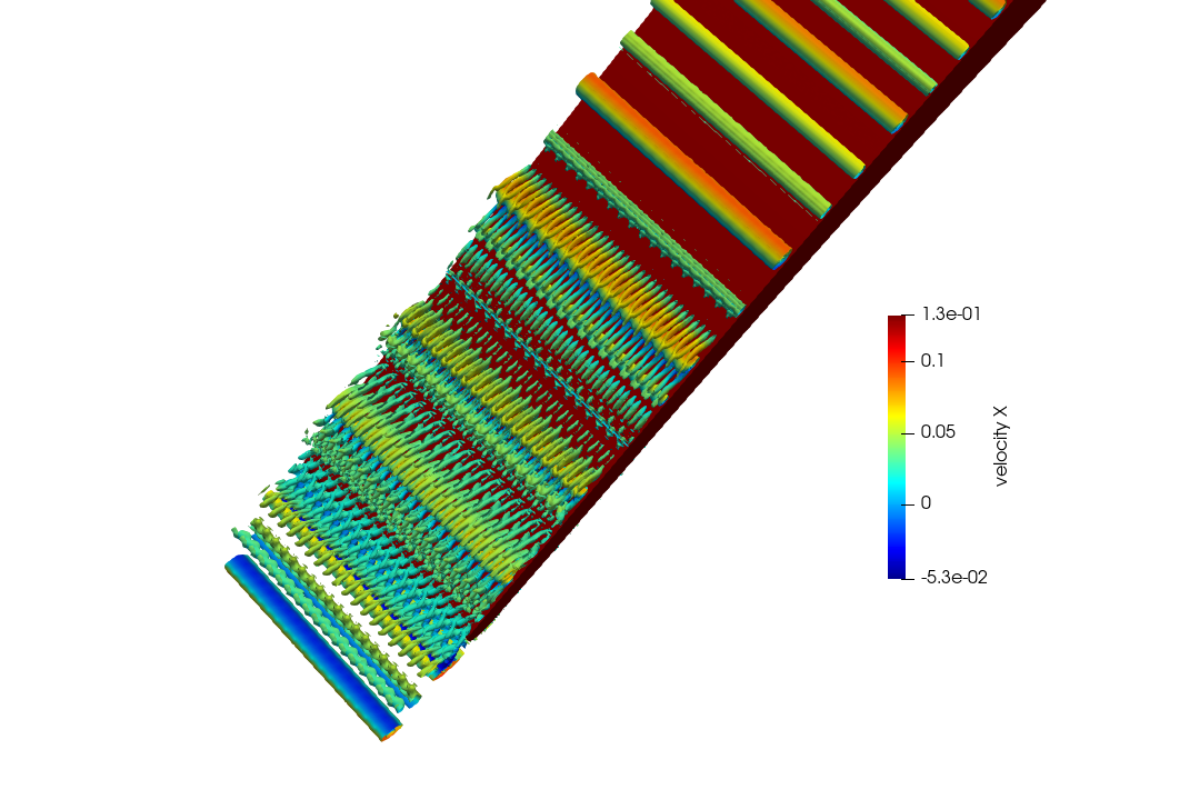}
		\caption{}
	\end{subfigure}
	\caption{Isosurfaces of Q-criterion coloured with stream-wise velocity for the inlet case }
	\label{fig:q_t106a_clean}
\end{figure}

\begin{figure}[h!]
	\centering
	\begin{subfigure}[b]{0.49\linewidth}
		\includegraphics[trim = 0cm 0cm 0cm 0cm, clip, width=\linewidth]{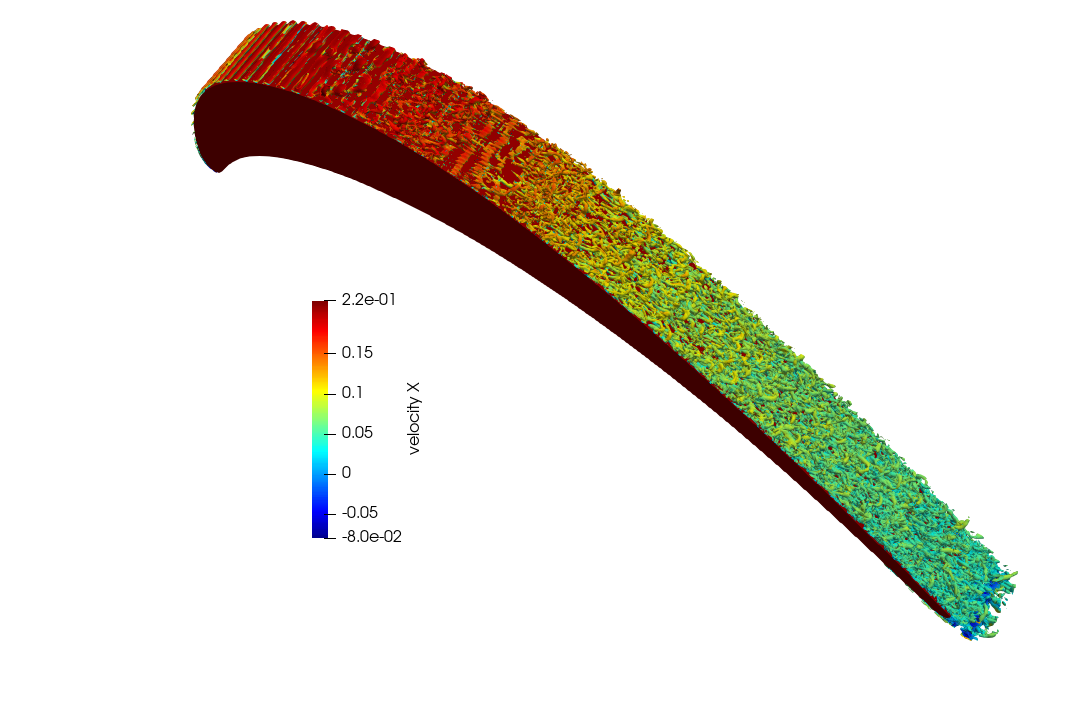}
		\caption{}
	\end{subfigure}
	\begin{subfigure}[b]{0.49\linewidth}
		\includegraphics[trim = 0cm 0cm 0cm 0cm, clip, width=\linewidth]{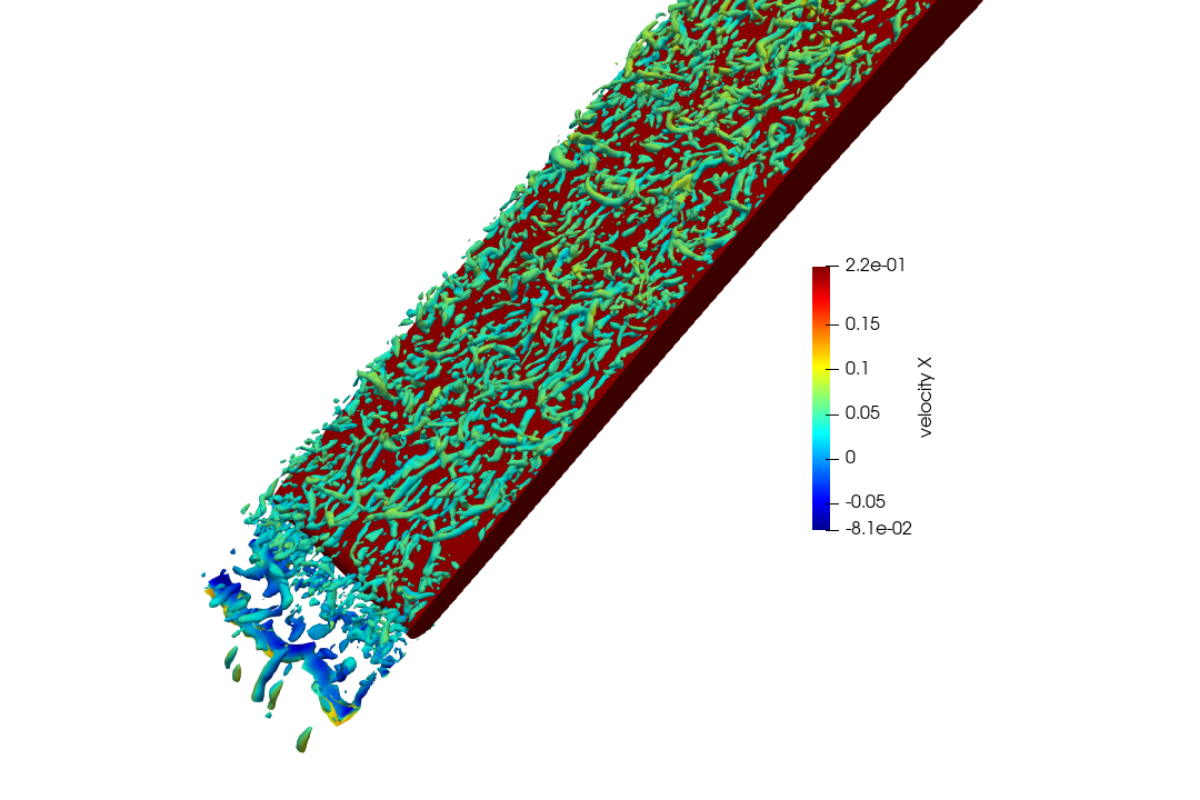}
		\caption{}
	\end{subfigure}
	\caption{Isosurfaces of Q-criterion coloured with stream-wise velocity for the inlet flow with noise }
	\label{fig:q_t106a_noise}
\end{figure}

Two sets of simulations with 1024 and 1536 grid points on the chord were carried out respectively, for both the cases of clean inlet and inlet with disturbance.
In Fig. \ref{fig:cp_t106a}, the time-averaged coefficient of pressure ($C_p = \frac{2(p-p_{ref})}{\rho U^2}$) is shown on the blade surface. The reference pressure is the inlet pressure averaged over the pitch. As can be seen from the figures, the $C_p$ result with clean inlet matches well with the experimental results, whereas the result with inlet disturbance overpredicts surface pressure near the trailing edge. Inlet disturbance has no effect on the pressure side.
Fig. \ref{fig:inst_mag} shows the contours of the magnitude of instantaneous velocity compared with the reference solution \cite{kalitzin2003dns}. The reference solution is obtained with inlet turbulence, which induces bypass transition. In the turbulence-free inlet case, ideally the flow undergoes a natural transition from laminar to turbulent flow due to the instabilities arising at the separation bubble on the suction side. However, in the present case, we do not observe bypass transition but the flow does become turbulent on the suction side.
To visualize the transition, we have plotted the y-component of velocity on the blade surface in Fig. \ref{fig:trans} along with the reference solution in which the velocity component normal to the blade surface is shown (Fig. \ref{fig:trans}c). Fig. \ref{fig:trans}a is obtained with clean inlet, and Fig. \ref{fig:trans}b is obtained with inlet disturbance. These figures clearly show the effect of inflow noise in tripping the flow to turbulence as inflow disturbances penetrate the boundary layer and trigger the separated shear layer.
For better visualization of flow structures near the wall, isosurfaces of instantaneous Q-criterion are plotted in Fig. \ref{fig:q_t106a_clean} and \ref{fig:q_t106a_noise} for clean inlet and for inflow with noise respectively. As can be seen from these figures, for the clean inlet, the flow remains laminar throughout the blade surface and ordered $\Lambda$ (Lambda) type vortices can be seen near the trailing edge. For the inlet case with noise, $\Lambda$ structures form hairpin vortices downstream and further break down to form small scale structures. Although the flow structures near the blade surface are drastically different for the two cases, the pressure coefficient is similar as evidenced in Fig. \ref{fig:cp_t106a}. The comparisons given herein are for the qualitative flow structures (see Fig. \ref{fig:trans}, \ref{fig:q_t106a_clean} and \ref{fig:q_t106a_noise}), more detailed quantitative analysis of the flow field is in progress.
\section{T106C}
\begin{figure}[h!]
	\centering
	\begin{subfigure}[b]{0.49\linewidth}
		\includegraphics[width=\linewidth]{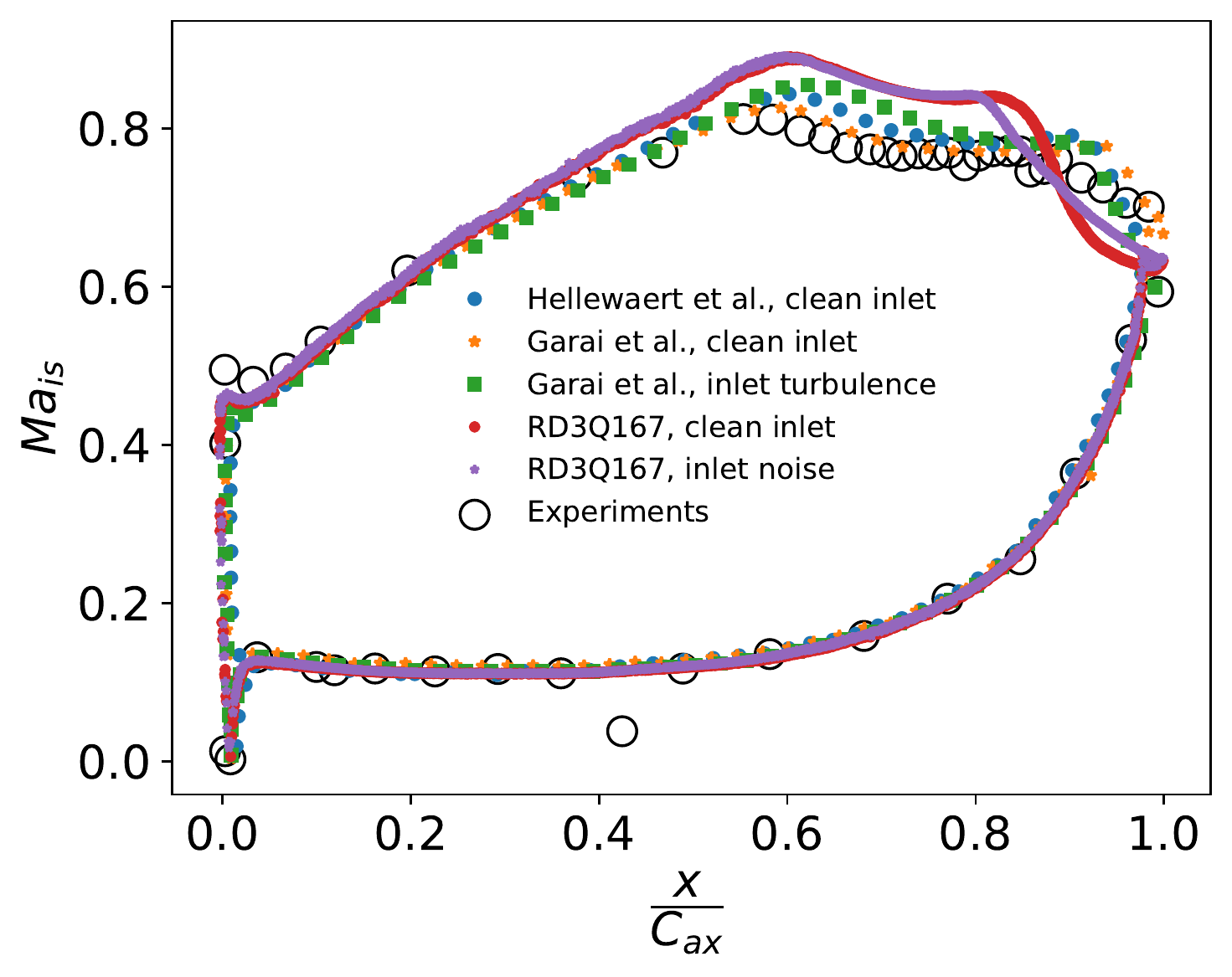}
		\caption{}
		\label{}
	\end{subfigure}
	\begin{subfigure}[b]{0.49\linewidth}
		\includegraphics[width=\linewidth]{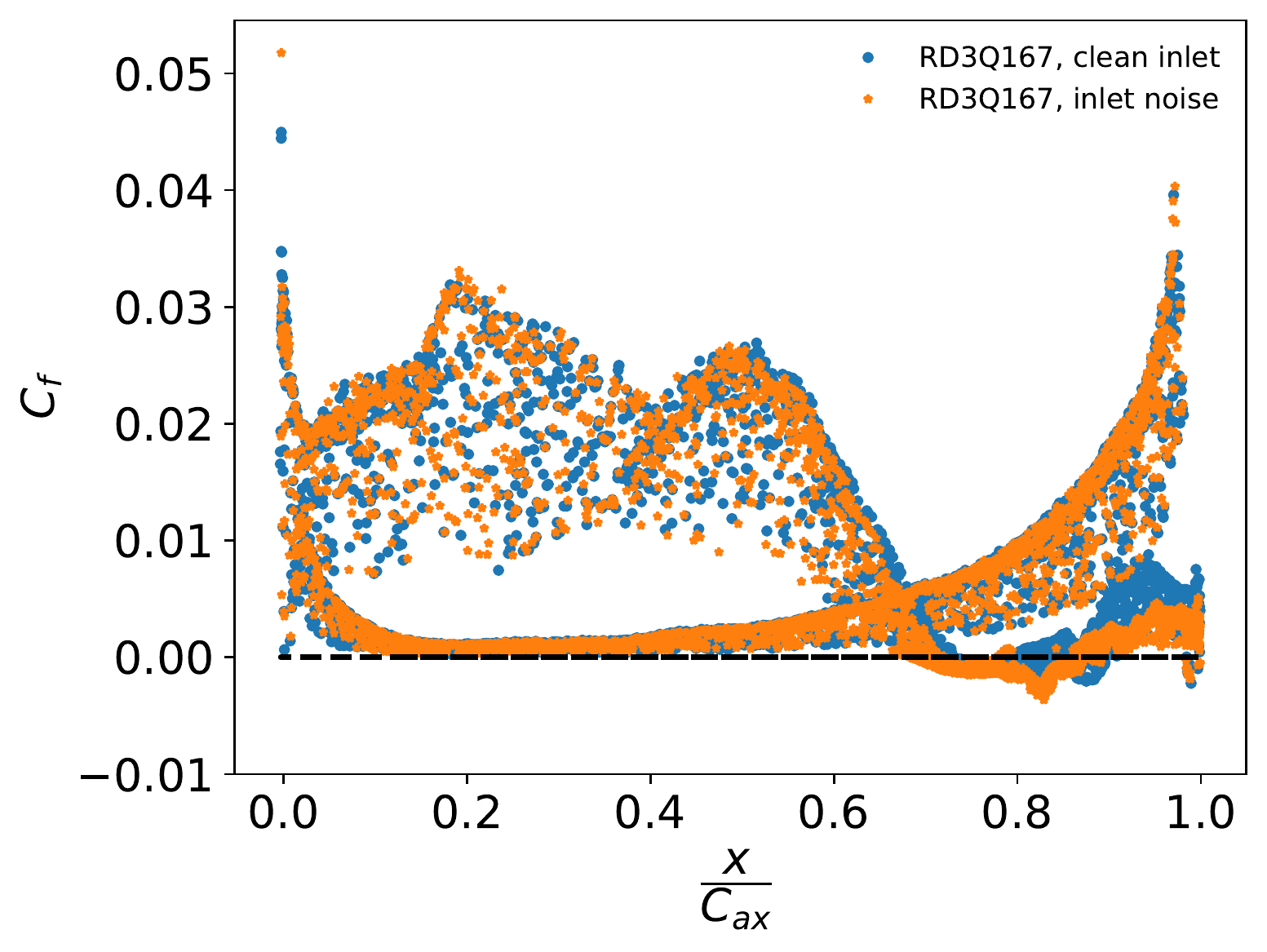}
		\caption{}
	\end{subfigure}
	\caption{Isentropic Mach number (a) and skin-friction coefficient (b) on the surface of the blade}
	\label{fig:t106c_cf}
\end{figure}

\begin{figure}[h!]
	\centering
	\includegraphics[width=0.75\linewidth]{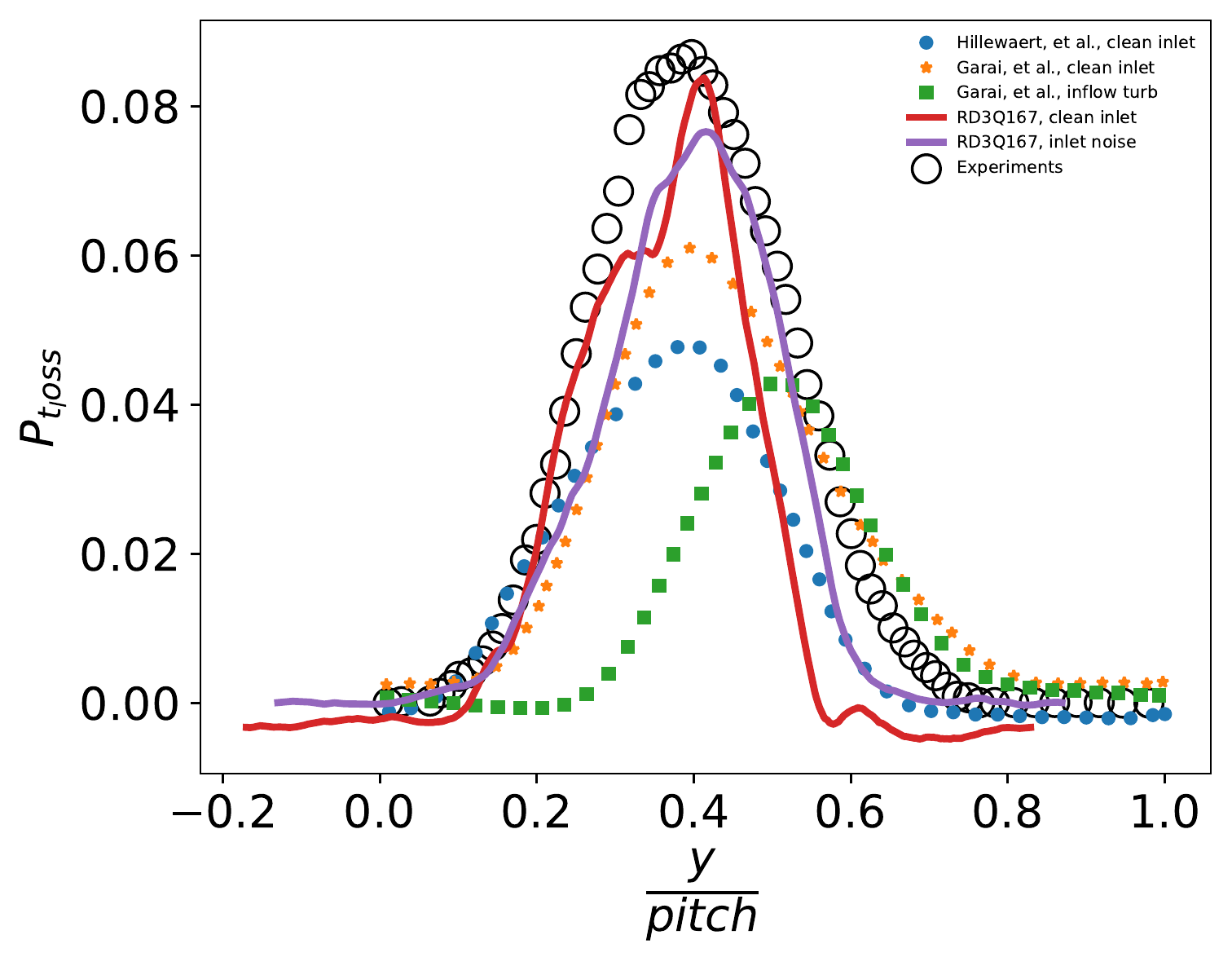}
	\caption{Total pressure loss in the wake of the blade}
	\label{fig:wake_loss}
\end{figure}

\begin{figure}[h!]
	\centering
	\begin{subfigure}[b]{0.49\linewidth}
		\includegraphics[width=\linewidth]{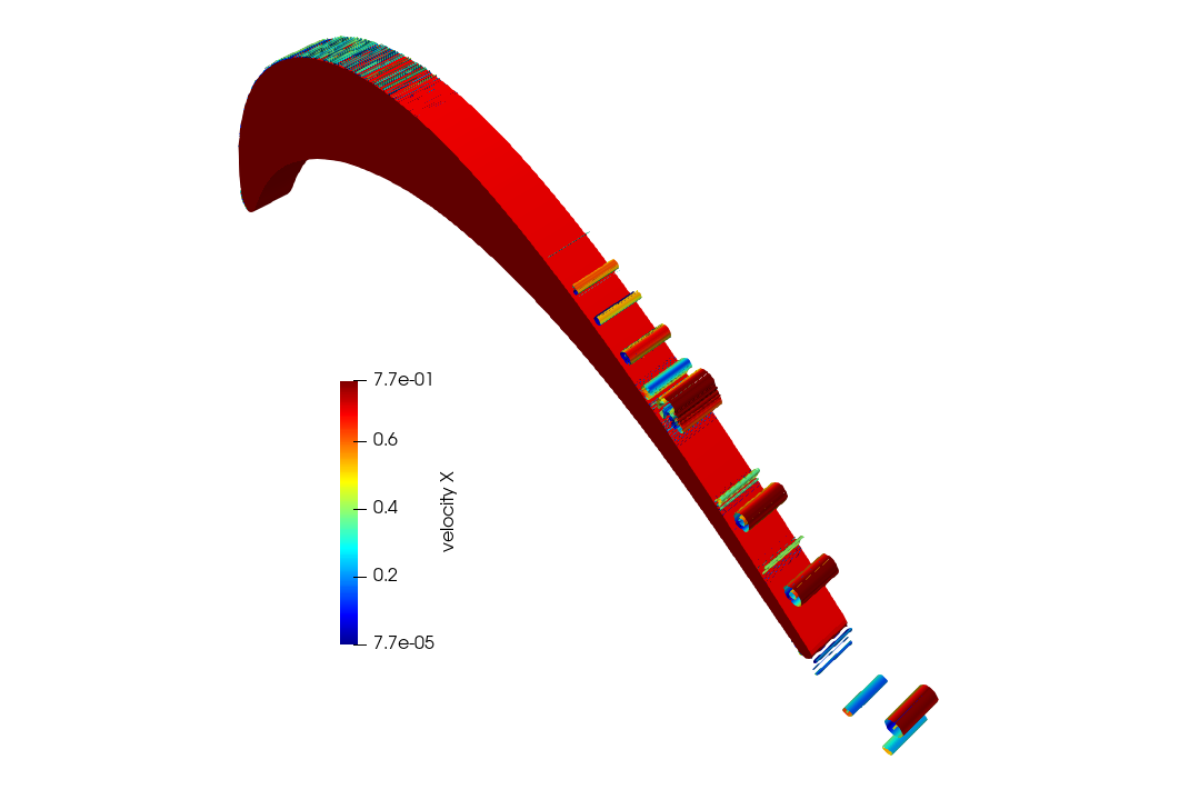}
		\caption{}
	\end{subfigure}
	\begin{subfigure}[b]{0.49\linewidth}
		\includegraphics[width=\linewidth]{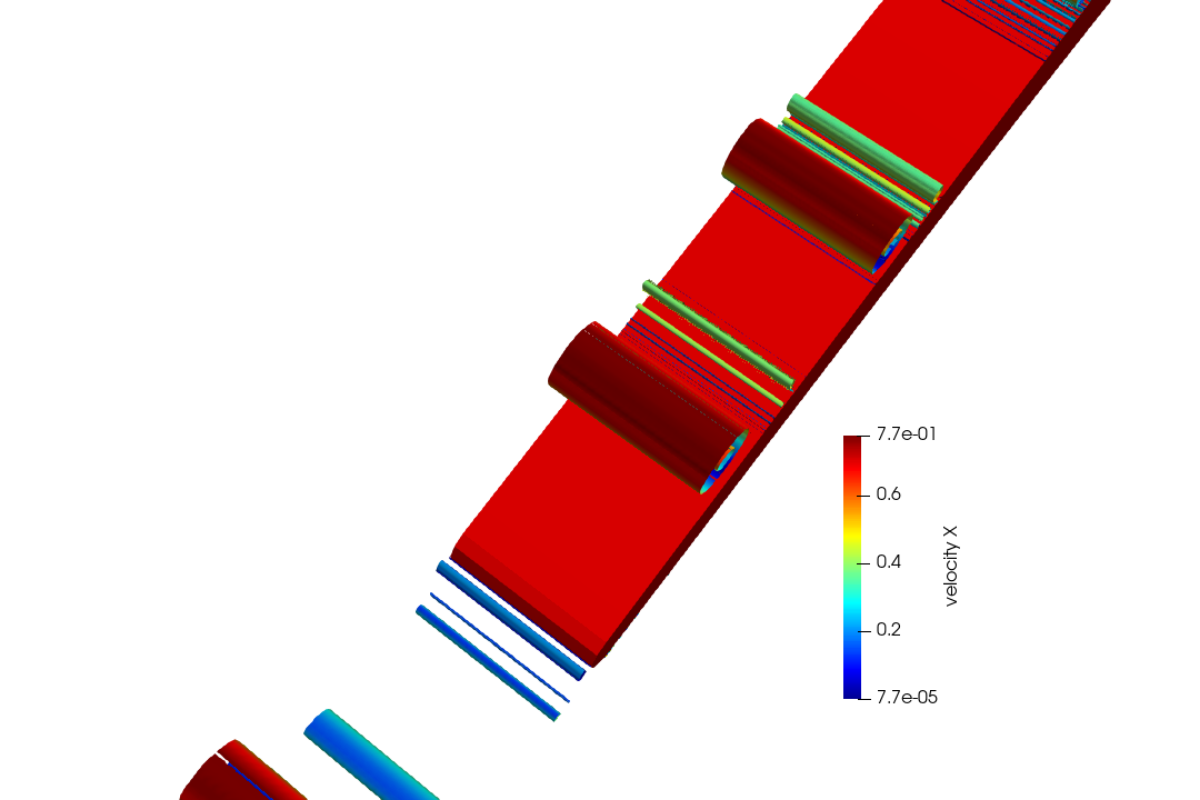}
		\caption{}
	\end{subfigure}
	\caption{Isosurfaces of Q (>0) criterion coloured with stream-wise velocity obtained with clean inlet flow}
	\label{fig:t106c_q_clean}
\end{figure}

\begin{figure}[h!]
	\centering
	\begin{subfigure}[b]{0.49\linewidth}
		\includegraphics[width=\linewidth]{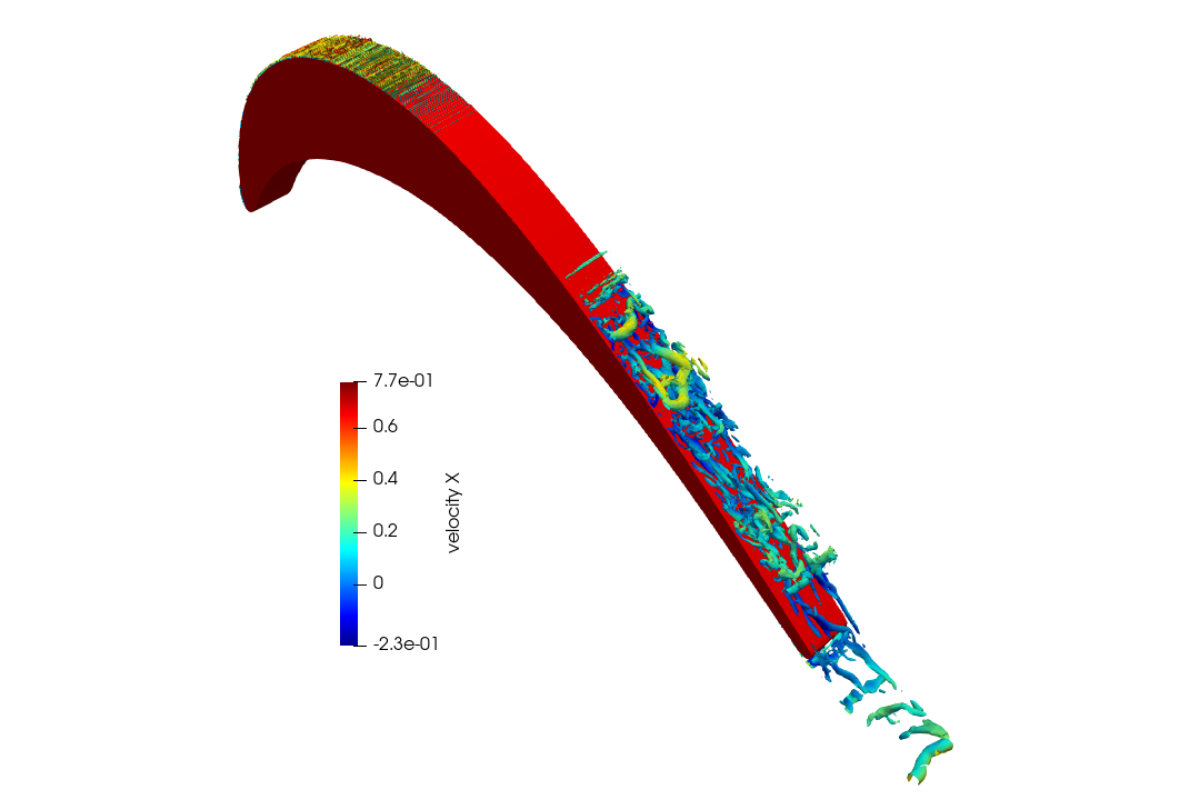}
		\caption{}
		\label{fig:Ma_is}
	\end{subfigure}
	\begin{subfigure}[b]{0.49\linewidth}
		\includegraphics[width=\linewidth]{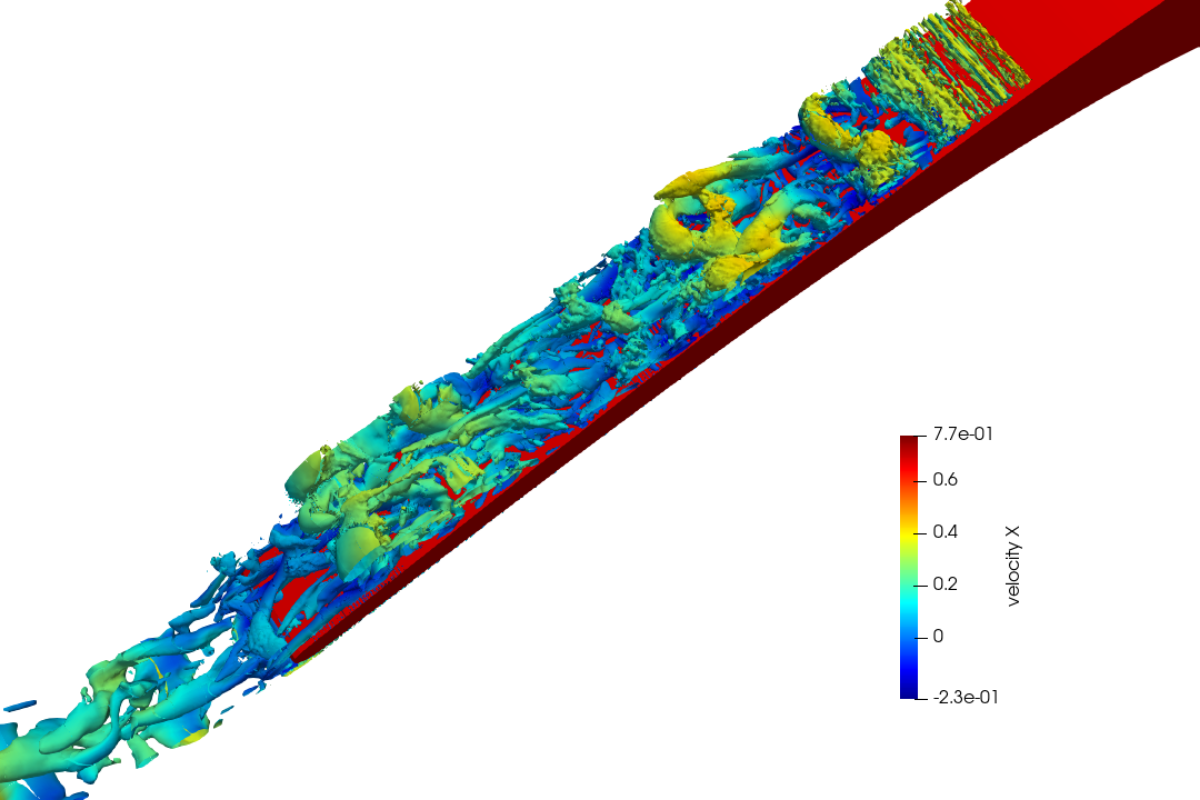}
		\caption{}
	\end{subfigure}
	\caption{Isosurfaces of Q (>0) criterion coloured with stream-wise velocity obtained with inflow disturbance}
	\label{fig:t106c_q_noise}
\end{figure}

Both clean inlet and inlet disturbance cases are simulated with 2048 points on the axial chord. Fig. \ref{fig:Ma_is} shows the isentropic Mach number on the airfoil surface in comparison with experimental results. As can be seen from the figure, the computed isentropic Mach number exactly matches with experimental results on the pressure side, whereas it is over predicted on the suction side near the trailing edge. This result is consistent with the earlier computational results reported in the literature. This maybe due to the discrepancy between experimental and computational setups, mainly the inflow and stagger angle of the blades \cite{hillewaert2014assessment}. We can observe that the inlet disturbance has a minor effect on the isentropic Mach number on the surface, specifically towards the trailing edge on the suction side. Our results are closer to Garai et al \cite{garai2016dns}., who carried out simulations with inlet turbulence of $3.2\%$.
Fig.\ref{fig:t106c_cf} shows the skin friction coefficient. Effect of inlet disturbance can be seen on the suction side towards the trailing edge. Due to the inlet disturbance, location of the flow separation changes. The flow remains attached on the pressure side for both the clean inlet and inlet with disturbances. For the clean inlet case, flow separates at $0.7C_{ax}$ and reattaches at $0.89C_{ax}$. For the inlet disturbance case, flow separates at $0.68C_{ax}$ and reattaches at $0.89C_{ax}$. However, Garai et al \cite{garai2016dns}., reported a separation point of $0.67C_{ax}$ and $0.74C_{ax}$ for the clean inlet and inlet turbulence case respectively with $0.96C_{ax}$ as a reattachment point for inlet turbulence case.
Fig. \ref{fig:wake_loss} shows the wake loss profiles in comparison with experimental and simulation results reported in the literature. Pressure loss profiles are extracted in the wake at $x=0.465C_{ax}$. As can be seen from figures, wake loss profiles reported in the literature widely vary with respect to experiments. One reason could be the discrepancy reported in the surface pressure. Present results with clean inlet predicts the peak loss similar to that of experiments. We have shifted the curve by $17\%$ of the pitch to match peak and width of the wake loss measured from experiments. However, pressure loss is under predicted for the inlet disturbance case for which the profile has been shifted by $12\%$ to match the peak. This result is similar to that of Mitra et al \cite{pratik2018les}., who have shifted the curve by $10\%$ and reported the peak loss of 0.05 for both clean inlet and inlet turbulence case.
To understand the transition process we have plotted Q criterion coloured with stream-wise velocity (see Fig. \ref{fig:t106c_q_clean} and \ref{fig:t106c_q_noise}). As can be seen from figures, flow remains attached and laminar until about $60$ to $70\%$ of the blade, beyond which the flow separates due to adverse pressure gradient although remaining laminar in the case of clean inlet. However, for the inlet disturbance case, separated shear layer undergoes transition to turbulence (due to the disturbances triggering the boundary layer), which can be clearly seen from isosurfaes of the Q-criterion. The flow on the pressure side is laminar and attached for both the cases.
\section{Summary}
Preliminary results on the incompressible and compressible flow past the T106 cascade have been reported. We have carried out simulations for both cases of clean inlet and inlet with disturbances. It has been shown that the pressure coefficient (for T106A) and isentropic Mach number (for T106C) on the surface of the blades matches well with the experimental results. For both the turbine blades with clean inlet, the flow remains laminar. However, with inlet disturbances, the flow undergoes transition to turbulence. Isosurfaces of Q-criterion show the transition zone for the inlet disturbance cases, which qualitatively agree with the reference solutions. We have observed that with clean inlet, the flow may not transition to turbulence. For the T106C cascade, peak wake loss matches well with experimental data and for the case of clean inlet while for the case of inlet disturbance, results are consistent with those reported in the literature. Further analysis is being carried out to study the transition and separation mechanisms over these blades.
\bibliography{sample}

\begin{thebibliography}{20}
\newcommand{\enquote}[1]{``#1''}
\providecommand{\natexlab}[1]{#1}
\providecommand{\url}[1]{\texttt{#1}}
\providecommand{\urlprefix}{URL }
\expandafter\ifx\csname urlstyle\endcsname\relax
  \providecommand{\doi}[1]{\discretionary{}{}{}https://doi.org/#1}\else
  \providecommand{\doi}[1]{\discretionary{}{}{}\urlstyle{rm}\url{https://doi.org/#1}}\fi

\bibitem[{Succi(2015)}]{succi2015lattice}
Succi, S., \enquote{Lattice boltzmann 2038,} \emph{EPL (Europhysics Letters)},
  Vol. 109, No.~5, 2015, p. 50001.

\bibitem[{Succi(2001)}]{succi2001lattice}
Succi, S., \emph{The lattice Boltzmann equation: for fluid dynamics and
  beyond}, Oxford university press, 2001.

\bibitem[{Karlin et~al.(1999)Karlin, Ferrante, and
  {\"O}ttinger}]{karlin1999perfect}
Karlin, I.~V., Ferrante, A., and {\"O}ttinger, H.~C., \enquote{Perfect entropy
  functions of the lattice Boltzmann method,} \emph{EPL (Europhysics Letters)},
  Vol.~47, No.~2, 1999, p. 182.

\bibitem[{Succi et~al.(2002)Succi, Karlin, and Chen}]{succi2002colloquium}
Succi, S., Karlin, I.~V., and Chen, H., \enquote{Colloquium: Role of the H
  theorem in lattice Boltzmann hydrodynamic simulations,} \emph{Reviews of
  Modern Physics}, Vol.~74, No.~4, 2002, p. 1203.

\bibitem[{Karlin et~al.(1998)Karlin, Gorban, Succi, and
  Boffi}]{karlin1998maximum}
Karlin, I.~V., Gorban, A.~N., Succi, S., and Boffi, V., \enquote{Maximum
  entropy principle for lattice kinetic equations,} \emph{Physical Review
  Letters}, Vol.~81, No.~1, 1998, p.~6.

\bibitem[{Ansumali and Karlin(2002)}]{ansumali2002entropy}
Ansumali, S., and Karlin, I.~V., \enquote{Entropy function approach to the
  lattice Boltzmann method,} \emph{Journal of Statistical Physics}, Vol. 107,
  No.~1, 2002, pp. 291--308.

\bibitem[{Ansumali et~al.(2003)Ansumali, Karlin, and
  {\"O}ttinger}]{ansumali2003minimal}
Ansumali, S., Karlin, I.~V., and {\"O}ttinger, H.~C., \enquote{Minimal entropic
  kinetic models for hydrodynamics,} \emph{EPL (Europhysics Letters)}, Vol.~63,
  No.~6, 2003, p. 798.

\bibitem[{Chikatamarla et~al.(2006)Chikatamarla, Ansumali, and
  Karlin}]{chikatamarla2006entropic}
Chikatamarla, S., Ansumali, S., and Karlin, I.~V., \enquote{Entropic lattice
  Boltzmann models for hydrodynamics in three dimensions,} \emph{Physical
  review letters}, Vol.~97, No.~1, 2006, p. 010201.

\bibitem[{Atif et~al.(2017)Atif, Kolluru, Thantanapally, and
  Ansumali}]{atif2017essentially}
Atif, M., Kolluru, P.~K., Thantanapally, C., and Ansumali, S.,
  \enquote{Essentially entropic lattice Boltzmann model,} \emph{Physical review
  letters}, Vol. 119, No.~24, 2017, p. 240602.

\bibitem[{Namburi et~al.(2016)Namburi, Krithivasan, and
  Ansumali}]{namburi2016crystallographic}
Namburi, M., Krithivasan, S., and Ansumali, S., \enquote{Crystallographic
  lattice Boltzmann method,} \emph{Scientific reports}, Vol.~6, No.~1, 2016,
  pp. 1--10.

\bibitem[{Kolluru et~al.(2020)Kolluru, Atif, Namburi, and
  Ansumali}]{kolluru2020lattice}
Kolluru, P.~K., Atif, M., Namburi, M., and Ansumali, S., \enquote{Lattice
  Boltzmann model for weakly compressible flows,} \emph{Physical Review E},
  Vol. 101, No.~1, 2020, p. 013309.

\bibitem[{Atif et~al.(2018)Atif, Namburi, and Ansumali}]{atif2018higher}
Atif, M., Namburi, M., and Ansumali, S., \enquote{Higher-order lattice
  Boltzmann model for thermohydrodynamics,} \emph{Physical Review E}, Vol.~98,
  No.~5, 2018, p. 053311.

\bibitem[{Stieger(2002)}]{stieger2002effect}
Stieger, R.~D., \enquote{The effect of wakes on separating boundary layers in
  low pressure turbines,} Ph.D. thesis, University of Cambridge, 2002.

\bibitem[{Krithivasan et~al.(2014)Krithivasan, Wahal, and
  Ansumali}]{PhysRevE.89.033313}
Krithivasan, S., Wahal, S., and Ansumali, S., \enquote{Diffused bounce-back
  condition and refill algorithm for the lattice Boltzmann method,} \emph{Phys.
  Rev. E}, Vol.~89, 2014, p. 033313.
\newblock \doi{10.1103/PhysRevE.89.033313},
  \urlprefix\url{https://link.aps.org/doi/10.1103/PhysRevE.89.033313}.

\bibitem[{Kalitzin et~al.(2003)Kalitzin, Wu, and Durbin}]{kalitzin2003dns}
Kalitzin, G., Wu, X., and Durbin, P.~A., \enquote{DNS of fully turbulent flow
  in a LPT passage,} \emph{International Journal of Heat and Fluid Flow},
  Vol.~24, No.~4, 2003, pp. 636--644.

\bibitem[{Koen~Hillewaert(2017)}]{KoenHillewaert}
Koen~Hillewaert, J.-S.~C., \enquote{{Spanwise periodic DNS/LES of the
  transitional flow in T106 LP} turbine cascades,}
  \url{https://how5.cenaero.be/sites/how5.cenaero.be/files/CS2_DNSLEST106_0.pdf},
  2017.

\bibitem[{Michalek et~al.(2012)Michalek, Monaldi, and
  Arts}]{michakalek2012aerodynamic}
Michalek, J., Monaldi, M., and Arts, T., \enquote{Aerodynamic performance of a
  very high lift low pressure turbine airfoil (T106C) at low Reynolds and high
  Mach number with effect of free stream turbulence intensity,} \emph{Journal
  of Turbomachinery}, Vol. 134, No.~6, 2012, p. 061009.

\bibitem[{Hillewaert et~al.(2014)Hillewaert, Carton~de Wiart, Verheylewegen,
  and Arts}]{hillewaert2014assessment}
Hillewaert, K., Carton~de Wiart, C., Verheylewegen, G., and Arts, T.,
  \enquote{Assessment of a high-order discontinuous Galerkin method for the
  direct numerical simulation of transition at low-Reynolds number in the T106C
  high-lift low pressure turbine cascade,} \emph{Turbo Expo: Power for Land,
  Sea, and Air}, Vol. 45615, American Society of Mechanical Engineers, 2014, p.
  V02BT39A034.

\bibitem[{Garai et~al.(2016)Garai, Diosady, Murman, and Madavan}]{garai2016dns}
Garai, A., Diosady, L.~T., Murman, S.~M., and Madavan, N.~K., \enquote{DNS of
  low-pressure turbine cascade flows with elevated inflow turbulence using a
  discontinuous-Galerkin spectral-element method,} \emph{Turbo Expo: Power for
  Land, Sea, and Air}, Vol. 49712, American Society of Mechanical Engineers,
  2016, p. V02CT39A025.

\bibitem[{Mitra and Mathew(2018)}]{pratik2018les}
Mitra, P., and Mathew, J., \enquote{LARGE EDDY SIMULATION OF TURBINE
  AERODYNAMICS BY EXPLICIT FILTERING,} , 2018.

\end{thebibliography}
\end{document}